\documentstyle[12pt,aasms4,flushrt,epsfig]{article}
\long\def\***#1{{\scshape ***#1***}}



\lefthead{SONGAILA}
\righthead{IGM TRANSMISSION}

\begin{document}

%\submitted{\today\ \  Submitted to the Astronomical Journal}
\title{The Evolution of the Intergalactic Medium Transmission to Redshift Six}
\author{Antoinette Songaila\altaffilmark{1}}
\affil{Institute for Astronomy, University of Hawaii, 2680 Woodlawn Drive,
  Honolulu, HI 96822\\}

\altaffiltext{1}{Visiting astronomer, W. M. Keck
  Observatory, jointly operated by the California Institute of Technology and
  the University of California.}

%\slugcomment{Astronomical Journal, submitted}
\slugcomment{To be published in Astronomical Journal, May 2004}

\begin{abstract}

We have measured the transmission of the Ly$\alpha$\ forest produced
by neutral hydrogen scattering in the intergalactic medium between
redshifts 2 and 6.3 using high signal to noise, high resolution ($R
\ge 5000$) observations of 50 quasars spread over the redshift range.
We use a uniform set of $15~{\rm\AA}$\ intervals covering Ly$\alpha$,
Ly$\beta$, and Ly$\gamma$\ absorption regions to tabulate the forest
transmission as a function of redshift.  The transmitted fractions
show a relatively smooth evolution over the entire range of redshifts,
which can be modelled with a smoothly decreasing ionization rate.
Previous claims of an abrupt change at $z \sim 6$\ appear in part to be a
consequence of an incorrect conversion of Ly$\beta$\ to Ly$\alpha$\
optical depths.  The tabulated transmissions can be used to calculate
the colors of objects with a specified input spectrum as a function of
redshift.  We calculate the colors of a flat $f_{\nu}$\ galaxy with a
large intrinsic continuum break, as an important example.

\end{abstract}


\section{Introduction} \label{intro}

The epoch of hydrogen reionization is one of the landmarks of the high
redshift universe, providing constraints on the first light sources to
have formed and a crucial input to models of structure formation.
Recent dramatic progress has been made in determining limits on this
redshift, leading to an interesting observational situation.  On the
one hand, spectroscopy of the Lyman $\alpha$\ forest shows that the
intergalactic medium (IGM) is highly ionized at $z < 6$, becoming
progressively opaque to radiation blueward of the quasars' Ly$\alpha$\
emission, culminating in the spectrum of the $z = 6.28$\ QSO
SDSS~1030+0524 which exhibits a substantial dark region that is
consistent with no transmitted flux, i.e. a Gunn-Peterson trough at $z
= 6.05$\ (Becker et al.\ 2002, Songaila \& Cowie 2002; Fan et al.\
2002).  This high opacity has been used to infer that this redshift is
at the tail end of the hydrogen reionization epoch (Becker et al.\
2002; Djorgovski et al.\ 2002) and comparison with simulations suggest
the IGM could have been fully neutral at $z = 6.5$\ (Cen \& McDonald
2002; Gnedin 2002; Fan et al.\ 2002) though, as we point out in this
paper, some of the strongest claims in this regard are based on an
incorrect conversion of Ly$\beta$\ opacities to equivalent Ly$\alpha$\
opacities (Becker et al.\ 2002; White et al.\ 2003).  On the other
hand, the recent detection by WMAP of a large optical depth to
electron scattering (Bennett et al. 2003) is consistent with the
Universe having been reionized at $z \sim 15$\ and fully ionized
thereafter.

The exact significance of the Gunn-Peterson trough in SDSS 1030+0524
is still not clear.  It could follow from the ongoing thickening of
the Ly$\alpha$\ forest rather than representing the onset of
reionization since it is extremely difficult to distinguish the two
effects.  The discovery of three new $z > 6$\ SDSS quasars (Fan et
al.\ 2003) has made the situation more complex since, as might have
been expected, there is significant variance in opacity between
different lines of sight at $z \sim 6$ (Fan et al.\ 2003; White et
al.\ 2003).  Although White et al.\ (2003) have argued that apparent
increased transmission in the spectrum of SDSS 1148+5251, currently
the highest redshift quasar, could be a result of foreground emission,
it is more likely to be real cosmic variance (Songaila 2004a).  In
assessing the significance of the opacity at $z = 6$\ we need to know
how the ionization rate in the IGM is evolving at lower redshifts,
since the strongest signature of reionization would be an abrupt
change in ionization at some redshift.  In this paper, we use high
signal-to-noise spectra of a large sample of quasars, including all
the brightest $z > 4.5$\ quasars observable from the northern
hemisphere, to investigate the evolution of the transmission in the
IGM from $z = 2$\ to $z = 6.3$.  We find that the transmission is
smoothly evolving with redshift throughout the range.

\section{Observations} \label{obs}

Spectra of 25 high redshift quasars, with $4.17 < z_{\rm em} < 6.39$,
were obtained for this program using ESI (Sheinis et al.\ 2000) on the
KeckII telescope in echellette mode.  The resolution was $\sim 5300$\
for the $0.75^{\prime\prime}$\ slit width used and the wavelength
coverage is complete from $4000~{\rm \AA}$\ to $10,000~{\rm \AA}$.
Sample spectra and a more extensive discussion can be found in
Songaila \& Cowie (2002) and a complete catalog will be published in
Songaila (2004b).  At the red wavelengths, extremely high precision
sky subtraction is required since the sky lines can be more than two
orders of magnitude brighter than the
quasars. In order to deal with
this
issue, individual half-hour exposures were made, stepped along
the slit, and the median was used to provide the primary sky
subtraction. Total exposure times were around 3--7 hours per quasar,
and are summarized in Table~1a.  The frames
were then registered, filtered to remove cosmic rays and artifacts,
and then added. At this point a second linear fit to the slit profile
in the vicinity of the quasar was used to remove any small residual
sky. The observations were made at the parallactic angle and flux
calibrated using observations of white dwarf standards scattered
through the night. These standards were also used to remove telluric
absorption features in the spectra, including the various atmospheric
bands. The final extractions of the spectra were made using a
weighting provided by the profile of the white dwarf standards.
Measurements of residual flux in saturated regions of the spectra
suggest that the sky subtraction leaves about 1\% residuals in the
faintest quasars and smaller errors in the brightest ones.  We take
the larger value as our systematic error.  Wavelength calibrations
were obtained using third-order fits to CuAr and Xe lamp observations
at the begining and end of each night, and the absolute wavelength
scale was then obtained by registering these wavelength solutions to
the night sky lines. All wavelengths and redshifts are given in the
vacuum heliocentric frame.

A low redshift comparison was provided by spectra of 25 quasars
obtained for other programs.  This sample is summarized in Table~1b. 
These spectra were taken with the HIRES
spectrograph (Vogt et al.\ 1994) on the Keck~I telescope with slit width $1.1$~arcsec,
giving a resolution $R = 36,000$.  Details of the data reduction have
been described previously (Songaila 1998).

The simplest characterization of the effects of the scattering by the
forest is its averaged effect in depressing the continuum.  This is
usually characterized by the Oke indices (Oke \& Korycansky 1982) averaged
over the total Ly$\alpha$, Ly$\beta$, etc. regions of the spectrum, or
by the generalization to the transmitted fraction,
\begin{equation}
F \equiv \left \langle {{f_{\lambda}} \over {f_{\rm cont}}} \right \rangle
\end{equation}
\medskip\noindent
where $f_{\lambda}$\ is the observed spectrum, $f_{\rm cont}$\ is
the continuum level in the absence of scattering, and the brackets
indicate averaging over a specified wavelength region.

Two methods have been used to compute $f_{\rm cont}$.  For low
resolution spectra, or for high redshift spectra for which $F$\ is very
small, we can calculate $f_{\rm cont}$\ only by extrapolating from
emission line-free regions in the spectrum longward of the quasar's
Ly$\alpha$\ emission.  Ideally one fits both the normalization and the
power law index but in the highest redshift quasars there is seldom
enough line-free spectrum to justify the fitting of the power law
index, and we assume instead a fixed power law index (here $-1.25$)
and set the normalization using the region of the spectrum at a rest
wavelength of $1350~{\rm\AA}$.  When $F$\ is small, as is the case at
high redshifts, it is relatively insensitive to the choice of power law
extrapolation.  At lower redshifts, where $F$\ is close to 1, the
measurement is sensitive to the extrapolation, and here it is
preferable to make a local interpolation between the peaks in the
spectrum to determine the continuum (Zuo \& Lu 1993).  This in turn
depends on there being regions of low opacity where the spectrum is
close to the continuum value.

We have used the continuum interpolation procedure for all of the
HIRES data and for the ESI data at redshift less than $z = 4.5$, and
have used the extrapolated continuum procedure for the ESI data at
redshifts above 4.5.  We have compared the two methods in the redshift
range $4.0 < z < 4.5$\ where there are still enough low opacity
regions to allow us to use the continuum interpolation method but
where there is substantial absorption through most of the spectrum, so
that the transmission is only weakly dependent on the power law
extrapolation.  For the 85 measurements in this range in the ESI
spectra, we find that, for a power law extrapolation with index
$1.25$, the ratio of the measured value relative to the interpolated
continuum transmission is 0.84 with a dispersion of $\pm 0.18$.  For
an extrapolation with a power law index of 1.75, this becomes $0.92
\pm 0.20$.  The measured transmission fractions lie in the range 0.2
to 0.6 so this is a relatively small correction and the 
slightly lower value of the continuum interpolation method 
is caused in part by its tendency
to over-estimate the
transmission slightly because the continuum fitting regions have some
opacity themselves.  This effect becomes larger with increasing
redshift.  However, because the local continuum fitting method is more
robust for individual quasars, we have chosen to use it for the ESI
data below $z = 4.5$\ and it is these values that are quoted in
Table~2.

Fluxes were measured in regions of the spectrum away from the emission
lines, namely: for Ly$\alpha$, in 8 bins of width $15~{\rm\AA}$
between $1080~{\rm\AA}$\ and $1185~{\rm\AA}$; for Ly$\beta$, in two
$15~{\rm\AA}$\ bins from $980~{\rm\AA}$\ to $1010~{\rm\AA}$; for
Ly$\gamma$, in 1 bin of width $15~{\rm \AA}$\ from $955~{\rm\AA}$\ to
$970~{\rm\AA}$.  The mean transmitted fractions for Ly$\alpha$\ and
Ly$\beta$\ are plotted in Figs.~1a and 1b (small filled squares) as a
function of redshift for the $ z > 4.5$\ ESI sample.  Large open
squares indicate quasars with BALQSO signatures, but the transmission
of these objects do not appear to differ significantly at these
wavelengths and we do not subequently distinguish these.  The
transmissions as a function of redshift are summarized in Table~2.
The transmitted fluxes blueward of Ly$\alpha$\ and Ly$\beta$\ in the
HIRES sample, measured in the same rest-frame redshift bins as for the
ESI sample, are plotted as crosses in Figures~1a and 1b and tabulated
in Table~2.  Diamonds denote the ESI measurements with $4 < z < 4.5$,
made with the local interpolation method.  Values of the transmitted
flux in the redshift range $z < 5$\ are in general agreement with
the observations of Schneider, Schmidt \& Gunn (1991) and Kennefick,
Dgorgovski \& de Carvalho (1995) and with the earlier literature
quoted in these papers.  See also Songaila \& Cowie (2002)
for a comparison.  At redshifts $z > 5$, they are also in general
agreement with the results of White et al\ (2003).  The Ly$\alpha$\
optical depths from White et al.\ are shown in Fig.~2a for the quasars
SDSS~1148+5251 (open diamonds) and SDSS~1030+0524 (open squares).

For purposes of display, it is useful to define a quantity $\tau_{\rm
MEAN}$ = $-$log(transmitted fraction)but it is important to remember
that this is {\it not\/} the average of the optical depth.  In
particular, $\tau_{\rm MEAN}$\ for Ly$\beta$ is not simply related to
$\tau_{\rm MEAN}$\ for Ly$\alpha$\ but is a complex function of the
structure (Fan et al.\ 2002; Songaila \& Cowie 2002).  Incorrect
interpretation of this quantity as a simple optical depth, as has been
done in Becker et al.\ (2002) and White et al.\ (2003), results in large
overestimates of the Ly$\alpha$\ optical depth constraints based on
Ly$\beta$\ transmission.  Using the mean of the Ly$\alpha$\ and
Ly$\beta$\ transmissions, averaged in six-point bins, we have computed
the corresponding Ly$\alpha$\ and Ly$\beta$\ $\tau_{\rm MEAN}$\ as a
function of the median redshift of each bin.  These are shown in
Figures~2a and 2b, where the error bars show the range of optical
depths corresponding to the maximum and minimum values of transmitted
flux in each bin.

\section{Discussion} \label{disc}

A uniform ionized IGM would have a Ly$\alpha$\ optical depth
\begin{equation}
\tau_u = 14 \, \Gamma_{-12}^{-1}  \, T_4^{-0.75} \,
             \left ( {{\Omega_m} \over {0.35}} \right )^{-0.5}
             \left ( {{\Omega_bh^2} \over {0.0325}} \right )^2
             \left ( {{H_0} \over {65~{\rm km\ s^{-1}\ Mpc^{-1}}}} \right )^{-1}
             \left ( {{1+z} \over {7}} \right ) ^{4.5} 
       \equiv 14 g^{-1} \left ( {{1+z} \over{7}} \right )^{4.5}
\end{equation}
\medskip\noindent where $\Gamma_{-12}$\ is the local ionization rate
produced by the metagalactic ionizing flux in units of $10^{-12}~{\rm
s}^{-1}$, $T_4$\ is the gas temperature in units of $10^4$~K, and the
second part of the equation defines $g$, which is the normalized
ionization rate.  For such a medium, $\tau_{\beta}$\ is simply $= 0.16
\tau_{\alpha}$, where the numerical factor is the ratio of
($f\lambda$) for the two lines, where $f$\ is the oscillator strength.
(Note that Becker et al.\ (2002), Fan et al.\ (2002) and White et al.\
(2003) incorrectly used only the ratio of the oscillator strengths.)
The dashed lines in Figs. 2a and 2b show this shape for a constant
$\Gamma$\ and $T_4$ (or constant $g$), where we have chosen the
normalization to provide a rough match to the Ly$\alpha$\ lines.  This
greatly oversimplified model gives a surprisingly good representation
of the Ly$\alpha$\ transmission over the full redshift range but
underpredicts Ly$\beta$, as can be seen in Figure~2b.

At $z \sim 4.5$ the average-density IGM begins to have substantial
optical depth and it is at this point that very few regions return
close to the extrapolated power law continuum as even the underdense
regions have some significant opacity.  It is also at this point that
the transmission starts to fall extremely rapidly as the underdense voids
themselves become opaque and there are no longer any regions of the
IGM that transmit significantly.  In this limit, it is clear why the
ionized Gunn-Peterson formula is a relatively good approximation, since
the bulk of the volume of the IGM is opaque and also relatively slowly
evolving in filling factor and density.
Because the transmission at high redshifts is dominated by the lowest
density ``voids'' corresponding to the most underdense regions that
occupy a significant volume, and since these regions evolve in a very
slow and straightforward way, it is possible to characterize very
simply the evolution of the mean transmission.  Songaila \& Cowie
(2002) derived the formula 
\begin{equation}
F(z,g) = 4.5 \, g^{-0.28} \, 
         \left ( {{1+z} \over {7}} \right )^{2.2} \,
         \exp \left ( -4.4 \, g^{-0.4} \, 
                      \left ( {{1+z} \over {7}} \right )^3 \right )
\end{equation}\smallskip\noindent
for the mean Ly$\alpha$\ transmission as a function of redshift for
the $\Lambda$CDM model of Cen \& McDonald.  
This analytic result closely matches the exact calulations of of
McDonald and collaborators (McDonald \& Miralda-Escud\`e 2001, Cen \&
McDonald, 2002).  It can be generalized to calculate
the transmission at Ly$\beta$:
\begin{equation}
F_{\beta}(z,g) = F(g,z_{\beta}) \, F(g/0.16,z)
\end{equation}\medskip\noindent
which is simply the product of the Ly$\beta$\ transmission and Ly$\alpha$\
transmission at the redshift $1 + z_{\beta} \equiv (\lambda_{\beta} /
\lambda_{\alpha})(1+z)$.  
Since we do not expect any substantial dependence on cosmology
for any model that matches the lower redshift Ly$\alpha$\ forest,
these equations allow us to express the evolution of the transmission
in terms of the redshift evolution of the ionization parameter, $g$.
For Ly$\alpha$, we use Equation~(1) to fit the
transmitted flux, assuming $g$\ is a power law function of $(1+z)$,
and obtain 
%$g = 1.5 \times ((1+z)/5)^{-4.5}$.  
$g = 0.74 \times ((1+z)/6)^{-4.1}$.  
Following
the prescription given above, the Ly$\beta$\ transmission is 
then determined and is given by
equations (3) and (4).  For this more realistic
model, the ratio of the Ly$\beta$\ and Ly $\alpha$\ transmissions is
0.4, which is a factor of 2.5 times larger than the 
value of 0.16 which is valid for the ionized Gunn-Peterson
case.  As a consequence, models that fit Ly$\alpha$, such as the
simple power law ionization we have assumed here, also provide a
reasonable fit to Ly$\beta$, as can be seen in Figs. 2a and 2b.

However, rather than fitting simple models for the ionization, it is
better to simply invert the equation (McDonald et al.\ (2000)), and in
Figure~3 we have shown the ionization parameter at each redshift which
would be required to produce the observed mean transmission.  Using
Equation~(3), we have calculated the ionization rate corresponding to
the mean Ly$\alpha$\ transmissions computed in 12-point bins, starting
with the highest redshift.  These are displayed as the small filled
squares in Figure~3.  The horizontal error bars show the range of
redshift in each bin and the vertical error bars show the ionization
parameter corresponding to $\pm 1~\sigma$\ in the mean transmission in
each bin.  The open squares are the measurements of McDonald et al.\
(2000) at lower redshift.  The dashed line is the power law fit to the
presently measured values of $g$\ for $z > 4$\ given above.  The
simple formula of Equation~(2) is valid only at $z > 4$, so we show
the interpretation of the data only above that point.  As long as 
Ly~$\alpha$\ transmission is detectable, the Ly~$\alpha$\ measurement
should give the most sensitive and reliable measurement of $g$.
However where Ly$\alpha$\ is not seen, the Ly$\beta$\ can give a more
sensitive constraint because of its weaker oscillator strength (Becker
et al.\ 2002; Fan et al.\ 2002; White et al.\ 2003).  We note that there
is some sign in Fig.~2b that the Ly$\beta$\ opacity may be high
relative to Ly$\alpha$\ in the highest redshift bin.  The value of
$g$\ would be 0.03 at $z = 6.05$\ if we used Ly$\beta$, which is more
consistent with Fan et al.\ (2002) though it still lies above their
upper limit.  However, we emphasise that the Ly$\alpha$\ values are
more robust.  This drop off in the ionization parameter must also be
combined with the mean free path for ionizing photons to determine the
actual production rate of ionizing photons.  This in turn means that
the fall off in the rate of ionizing photon production is shallower
than the fall-off in $g$.  We postpone a more detailed discussion of
this to a subsequent paper.

\section{High Redshift Colors}\label{colors}

  Precise measurements of the forest transmission are also invaluable
for determining the colors of high redshift galaxies and
quasars. Currently the effects are usually included using the Madau
(1995) models which are based on early data in the range $2 < z < 5$.
The present data can be used instead to compute these colors on an
empirical basis and also to compute how much change can be produced by
cosmic variance in the forest scattering.  For broad band filters this
latter effect is very small at all redshifts but it can be significant
if narrow band filters are used.

As a useful example, we have computed the expected magnitude change
as a function of redshift for the case of a flat $f_{\nu}$\ spectrum
galaxy with a complete break at the Lyman continuum edge, which should
be representative of typical high redshift galaxies. For each of 5
colors (Johnson $B$\ and $V$, Kron-Cousins $R$\ and $I$\ and Sloan
$z$) we computed the convolution of the spectrum with the filter
transmission and the forest transmission and computed the decrement
relative to a flat $f_{\nu}$\ object without a break. For regions of
the spectrum where the Ly$\alpha$\ forest would lie in the filter we
interpolated the numbers in Table~2a to determine the average
transmission, and similarly for the Ly$\beta$\ transmission from
Table~2b.  For wavelengths between the Lyman continuum break and the
Ly$\gamma$\ line we used the results of Table~2c.  The results are
relatively insensitive to this approximation of the higher order Lyman
series scattering.  The results are shown in Figure~4 and summarised
in Table~3 where the changes in magnitudes produced by the decrement
are given. The difference between the magnitude shifts in any two
filters gives the color of the galaxy at the redshift in the $AB$\
magnitude system of Oke. Therefore, at $z=5.7$, for example, the
$AB$\ $(R-z)$\ color is 2.19. Similar results may be computed for any
input spectrum using the transmissions in the tables.

\section{Conclusions} \label{conc}

We have measured the transmission of the Ly$\alpha$\ forest produced
by neutral hydrogen scattering in the intergalactic medium between
redshifts 2 and 6.3 using high signal to noise, high resolution ($R\ge
5000$) observations of 50 quasars spread over the redshift range.  The
transmitted fractions show a relatively smooth evolution over the
entire range of redshifts, which can be modelled with a smoothly
decreasing ionization rate.  We have used the tabulated transmissions
to calculate the colors of a flat-spectrum galaxy with a large
intrinsic Lyman continuum break.

\acknowledgments 

I would like to thank X. Fan and R. Becker for providing the
coordinates of some of the $z > 6$\ quasars prior to their publication
and for useful comments on the first draft, and Michael Strauss and
Robert Lupton for helpful dicsussions.  Peter Shaver kindly provided a
critical reading of the first draft.  This research was supported by
the National Science Foundation under grant AST 00-98480.

%  TABLE 1a
\begin{deluxetable}{lccc}
\tablewidth{250pt}
\tablenum{1a}
\tablecaption{High Redshift ESI Observations \label{tbl:1a}}
\tablehead{
\colhead{Quasar} & \colhead{Mag.} & \colhead{Expo. (hrs)} 
& \colhead{$z_{em}$} } 
\startdata
SDSS 0206+1216    & 19.9  & 3.0  & 4.81  \nl
SDSS 0211$-$0009  & 20.0  & 7.0  & 4.90  \nl
SDSS 0231$-$0728  & 19.2  & 4.33 & 5.42  \nl
SDSS 0338+0021    & 20.0  & 7.25 & 5.01  \nl
BR 0353$-$3820    & 18.0  & 2.5  & 4.58  \nl
PSS 0747+4434     & 18.1  & 3.0  & 4.39?  \nl
FIRST 0747+2739   & 17.2  & 1.5  & 4.17  \nl
SDSS 0756+4104    & 20.1  & 3.0  & 5.09  \nl
SDSS 0836+0054    & 18.8  & 6.0  & 5.82  \nl
BRI 0952$-$0115   & 18.7  & 4.25 & 4.42  \nl
SDSS 1030+0524    & 19.7  & 6.28 & 6.28  \nl
BR 1033$-$0327    & 18.5  & 2.0  & 4.51  \nl
SDSS 1044$-$0125  & 19.7  & 5.75 & 5.74  \nl
SDSS 1048+4637    & 18.4  & 6.23 & 6.23  \nl
SDSS 1148+5251    & 19.1  & 6.39 & 6.39  \nl
SDSS 1204$-$0021  & 19.1  & 3.0  & 5.07  \nl
BR 1202$-$0725    & 18.7  & 3.0  & 4.61  \nl
SDSS 1306+0356    & 19.6  & 6.75 & 5.99  \nl
SDSS 1321+0038    & 20.1  & 2.0  & 4.71  \nl
SDSS 1605$-$0122  & 19.4  & 3.75 & 4.93  \nl
WFSJ 1612+5255    & 19.9  & 1.5  & 4.95  \nl
SDSS 1737+5828    & 19.3  & 7.5  & 4.85  \nl
SDSS 2200+0017    & 19.1  & 5.3  & 4.78  \nl
SDSS 2216+0013    & 20.3  & 3.0  & 5.00  \nl
BR 2237$-$0607    & 18.3  & 11.6 & 4.55  \nl

\enddata

%\tablenotetext{(a)}{Tablenotetext ...}

\end{deluxetable}

%  TABLE 1b
\begin{deluxetable}{lccc}
\tablewidth{250pt}
\tablenum{1b}
\tablecaption{Low Redshift HIRES Observations \label{tbl:1b}}
\tablehead{
\colhead{Quasar} & \colhead{Mag.} & \colhead{Expo. (hrs)} 
& \colhead{$z_{em}$} } 
\startdata
Q0014+813     & 16.5  & 7.8  &    3.38 \nl
HS0119+1432   & 16.7  & 7.8  &    2.87 \nl
Q0256-000     & 18.7  & 5.5  &    3.37 \nl
Q0302-003     & 18.4  & 10.7  &    3.28 \nl
Q0636+680     & 16.5  & 7.4  &    3.18 \nl
HS0741+4741   & 17.5  & 8.0  &    3.22 \nl
HS0757+5218   & 17.5  & 4.0  &    3.2 \nl
Q08279+5255   & 15.2  & 3.8  &    3.87 \nl
Q0956+122     & 18.2  & 7.3  &    3.30 \nl
Q1032+0414    & 18.1  & 4.0  &    3.39 \nl
J1057+4555    & 17.7  & 11.3  &    4.10 \nl
HE1104-1805   & 16.3  & 2.0  &    2.31 \nl
HE1122-1648   & 16.6  & 3.5  &    2.40 \nl
Q1159+123     & 17.5  & 8.0  &    3.50 \nl
Q1206+119     & 17.9  & 5.0  &    3.113 \nl
B1310+4254    & 18.4  & 5.2  &    2.561 \nl
Q1422+2309    & 16.7  & 16.2  &    3.63 \nl
Q1623+269     & 16.0  & 4.0  &    2.52 \nl
FBQS1634+3203  & 16.6  & 4.0  &    2.35 \nl
HS1700+6416   & 16.1  & 11.0  &    2.748 \nl
Q1937-1009    & 17.5  & 9.3  &    3.805 \nl
HS1946+7658   & 15.8  & 12.0  &    3.02 \nl
Q2000-330     & 17.6  & 4.7  &    3.77 \nl
Q2126-158     & 17.3  & 4.4  &    3.28 \nl
HE2347-4342   & 16.1  & 6.7  &    2.88 \nl
\enddata
 
%\tablenotetext{(a)}{Tablenotetext ...}
 
\end{deluxetable}

%  TABLE 2a
\begin{deluxetable}{lccc}
\tablewidth{250pt}
\tablenum{2a}
\tablecaption{Mean Transmission Blueward of Ly$\alpha$\label{tbl:2a}}
\tablehead{
\colhead{$z_{\rm median}$} 
& \colhead{$\langle T \rangle$} 
& \colhead{$T_{\rm max}$}  & \colhead{$T_{\rm min}$} } 
\startdata
2.40 & 0.820 & 0.900 & 0.710 \nl
2.46 & 0.809 & 0.879 & 0.735 \nl
2.54 & 0.818 & 0.899 & 0.761 \nl
2.61 & 0.778 & 0.833 & 0.700 \nl
2.67 & 0.728 & 0.839 & 0.608 \nl
2.72 & 0.800 & 0.893 & 0.702 \nl
2.75 & 0.780 & 0.837 & 0.700 \nl
2.79 & 0.764 & 0.811 & 0.648 \nl
2.83 & 0.698 & 0.806 & 0.575 \nl
2.86 & 0.695 & 0.806 & 0.552 \nl
2.89 & 0.729 & 0.808 & 0.643 \nl
2.92 & 0.694 & 0.799 & 0.492 \nl
2.95 & 0.788 & 0.847 & 0.744 \nl
2.98 & 0.712 & 0.812 & 0.644 \nl
3.01 & 0.723 & 0.833 & 0.618 \nl
3.03 & 0.692 & 0.821 & 0.456 \nl
3.08 & 0.697 & 0.774 & 0.639 \nl
3.09 & 0.719 & 0.855 & 0.545 \nl
3.14 & 0.767 & 0.849 & 0.711 \nl
3.19 & 0.717 & 0.785 & 0.650 \nl
3.24 & 0.675 & 0.713 & 0.630 \nl
3.35 & 0.687 & 0.800 & 0.597 \nl
3.48 & 0.628 & 0.730 & 0.532 \nl
3.59 & 0.597 & 0.729 & 0.467 \nl
3.65 & 0.553 & 0.633 & 0.449 \nl
3.75 & 0.495 & 0.574 & 0.414 \nl
3.84 & 0.495 & 0.624 & 0.352 \nl
3.89 & 0.563 & 0.803 & 0.290 \nl
3.94 & 0.487 & 0.612 & 0.363 \nl
3.99 & 0.467 & 0.565 & 0.346 \nl
4.03 & 0.388 & 0.469 & 0.302 \nl
4.08 & 0.434 & 0.577 & 0.297 \nl
4.12 & 0.491 & 0.640 & 0.367 \nl
4.17 & 0.487 & 0.554 & 0.419 \nl
4.20 & 0.441 & 0.577 & 0.342 \nl
4.23 & 0.414 & 0.492 & 0.348 \nl
4.26 & 0.404 & 0.487 & 0.309 \nl
4.29 & 0.300 & 0.453 & 0.135 \nl
4.31 & 0.439 & 0.588 & 0.326 \nl
4.35 & 0.350 & 0.420 & 0.279 \nl
4.37 & 0.398 & 0.551 & 0.219 \nl
4.41 & 0.426 & 0.621 & 0.288 \nl
4.44 & 0.348 & 0.397 & 0.253 \nl
4.48 & 0.328 & 0.401 & 0.277 \nl
4.52 & 0.316 & 0.408 & 0.187 \nl
4.56 & 0.376 & 0.472 & 0.247 \nl
4.59 & 0.297 & 0.366 & 0.148 \nl
4.66 & 0.222 & 0.317 & 0.110 \nl
4.68 & 0.215 & 0.273 & 0.093 \nl
4.74 & 0.238 & 0.303 & 0.194 \nl
4.82 & 0.246 & 0.334 & 0.200 \nl
4.98 & 0.182 & 0.222 & 0.151 \nl
5.13 & 0.147 & 0.321 & 0.095 \nl
5.26 & 0.084 & 0.208 & 0.016 \nl
5.42 & 0.090 & 0.150 & 0.027 \nl
5.51 & 0.060 & 0.161 & 0.025 \nl
5.60 & 0.082 & 0.178 & 0.044 \nl
5.70 & 0.028 & 0.071 & -0.001 \nl
5.87 & 0.009 & 0.025 & -0.001 \nl
6.05 & 0.0006 & 0.007 & -0.002 \nl

\enddata

%\tablenotetext{(a)}{Tablenotetext ...}

\end{deluxetable}

%  TABLE 2b
\begin{deluxetable}{lccc}
\tablewidth{250pt}
\tablenum{2b}
\tablecaption{Mean Transmission Blueward of Ly$\beta$\label{tbl:2b}}
\tablehead{
\colhead{$z_{\rm median}$} 
& \colhead{$\langle T \rangle$} 
& \colhead{$T_{\rm max}$}  & \colhead{$T_{\rm min}$} } 
\startdata
3.06 & 0.735 & 0.820 & 0.627 \nl
3.18 & 0.718 & 0.785 & 0.624 \nl
3.27 & 0.707 & 0.801 & 0.616 \nl
3.52 & 0.641 & 0.700 & 0.585 \nl
4.05 & 0.502 & 0.611 & 0.394 \nl
4.30 & 0.502 & 0.582 & 0.433 \nl
4.42 & 0.457 & 0.560 & 0.380 \nl
4.59 & 0.378 & 0.625 & 0.200 \nl
4.71 & 0.253 & 0.413 & -0.013 \nl
4.81 & 0.243 & 0.328 & 0.085 \nl
4.95 & 0.258 & 0.353 & 0.162 \nl
5.66 & 0.082 & 0.140 & 0.035 \nl
6.11 & 0.0002 & 0.009 & -0.008 \nl

\enddata

%\tablenotetext{(a)}{Tablenotetext ...}

\end{deluxetable}

%  TABLE 2c
\begin{deluxetable}{lccc}
\tablewidth{250pt}
\tablenum{2c}
\tablecaption{Mean Transmission Blueward of Ly$\gamma$\label{tbl:2c}}
\tablehead{
\colhead{$z_{\rm median}$} 
& \colhead{$\langle T \rangle$} 
& \colhead{$T_{\rm max}$}  & \colhead{$T_{\rm min}$} } 
\startdata
3.32 & 0.715 & 0.749 & 0.650 \nl
4.04 & 0.428 & 0.668 & -0.422 \nl
4.49 & 0.449 & 0.541 & 0.373 \nl
4.78 & 0.249 & 0.549 & -0.039 \nl
5.00 & 0.240 & 0.416 & 0.173 \nl
6.15 & 0.038 & 0.126 & -0.001 \nl

\enddata

%\tablenotetext{(a)}{Tablenotetext ...}

\end{deluxetable}

%  TABLE 3
\begin{deluxetable}{cccccc}
\tablewidth{250pt}
\tablenum{3}
\tablecaption{Colors of High redshift Galaxies and Quasars \label{tbl:3}}
\tablehead{
\colhead{Redshift} & \colhead{$z$} & \colhead{$I$} 
& \colhead{$R$} & \colhead{$V$} & \colhead{$B$} 
} 
\startdata
3.0 & 0.00 & 0.00 & 0.00 & 0.00 & 0.37  \nl
3.1 & 0.00 & 0.00 & 0.00 & 0.01 & 0.48  \nl
3.2 & 0.00 & 0.00 & 0.00 & 0.04 & 0.59  \nl
3.3 & 0.00 & 0.00 & 0.00 & 0.08 & 0.65  \nl
3.4 & 0.00 & 0.00 & 0.00 & 0.13 & 0.68  \nl
3.5 & 0.00 & 0.00 & 0.00 & 0.18 & 0.72  \nl
3.6 & 0.00 & 0.00 & 0.00 & 0.25 & 0.84  \nl
3.7 & 0.00 & 0.00 & 1.10 & 0.33 & 1.00  \nl
3.8 & 0.00 & 0.00 & 0.00 & 0.43 & 1.17  \nl
3.9 & 0.00 & 0.00 & 0.02 & 0.51 & 1.41  \nl
4.0 & 0.00 & 0.00 & 0.07 & 0.56 & 1.76  \nl
4.1 & 0.00 & 0.00 & 0.14 & 0.60 & 2.17  \nl
4.2 & 0.00 & 0.00 & 0.21 & 0.65 & 2.68  \nl
4.3 & 0.00 & 0.00 & 0.29 & 0.69 & 3.55  \nl
4.4 & 0.00 & 0.00 & 0.39 & 0.72 & 5.39  \nl
4.5 & 0.00 & 0.00 & 0.50 & 0.85 & 7.80  \nl
4.6 & 0.00 & 0.00 & 0.62 & 1.01 & 9.19  \nl
4.7 & 0.00 & 0.00 & 0.79 & 1.29 & 10.5  \nl
4.8 & 0.00 & 0.00 & 0.99 & 1.63 & 12.2  \nl
4.9 & 0.00 & 0.00 & 1.13 & 1.97 & 13.6  \nl
5.0 & 0.00 & 0.03 & 1.19 & 2.30 & 18.2  \nl
5.1 & 0.00 & 0.10 & 1.24 & 2.56 & 25.0  \nl
5.2 & 0.00 & 0.18 & 1.30 & 2.92 & 25.0  \nl
5.3 & 0.00 & 0.29 & 1.39 & 3.42 & 25.0  \nl
5.4 & 0.00 & 0.42 & 1.51 & 4.06 & 25.0  \nl
5.5 & 0.00 & 0.56 & 1.69 & 5.31 & 25.0  \nl
5.6 & 0.00 & 0.72 & 1.94 & 6.98 & 25.0  \nl
5.7 & 0.00 & 0.91 & 2.19 & 8.71 & 25.0  \nl
5.8 & 0.00 & 1.16 & 2.52 & 10.0 & 25.0  \nl
5.9 & 0.01 & 1.46 & 2.97 & 11.1 & 25.0  \nl
6.0 & 0.04 & 1.87 & 3.38 & 12.3 & 25.0  \nl
6.1 & 0.12 & 2.57 & 3.83 & 14.0 & 25.0  \nl
6.2 & 0.23 & 3.30 & 4.34 & 20.4 & 25.0  \nl
6.3 & 0.35 & 3.66 & 4.83 & 25.0 & 25.0  \nl
6.4 & 0.49 & 3.93 & 5.45 & 25.0 & 25.0  \nl
\enddata
\end{deluxetable}

\newpage
%
% Figure 1a
%
%\begin{inlinefigure}
\begin{figure}
\epsfig{figure=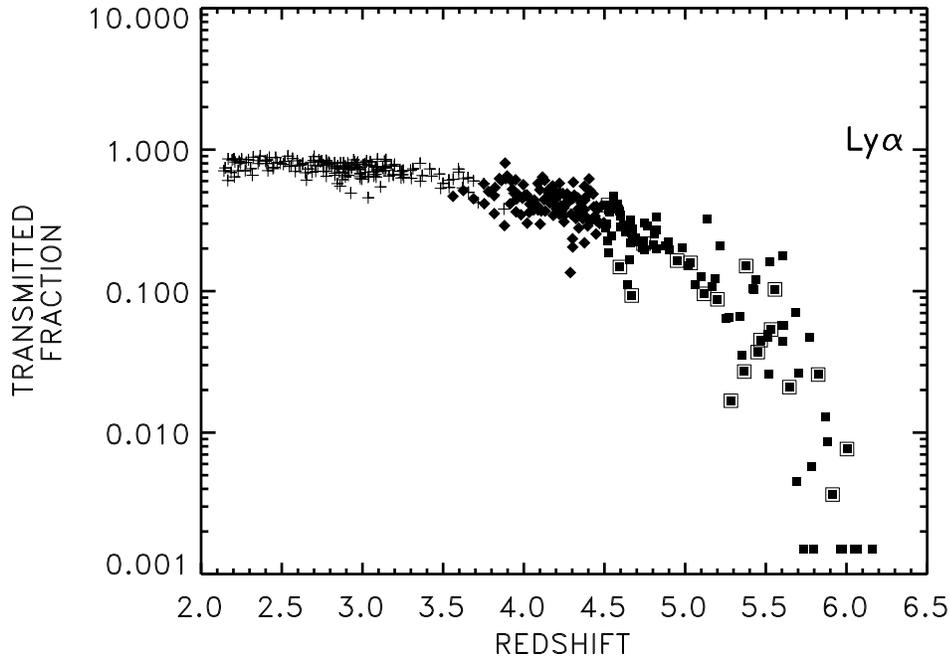,angle=0,width=5.0in}
\vspace{6pt}
\figurenum{1a}
\caption{
Transmitted flux blueward of Ly$\alpha$\
emission as a function of redshift for both the ESI and HIRES samples.
In all cases, flux was
computed in 8 bins of width $15~{\rm \AA}$\ between $1080~{\rm \AA}$\
and $1185~{\rm\AA}$. For the $z > 4.5$\ ESI sample (filled squares),
the flux was 
normalized to a quasar power law continuum
with slope $-1.25$\ normalized at $1350~{\rm\AA}$.  Large open squares
denote data for real or suspected BAL QSOs in the sample, namely
SDSS~1605$-$0112, SDSS~1044$-$0125, and SDSS~1048+4637.  For the lower
redshift HIRES sample (crosses), the flux was normalized to a continuum
that was interpolated locally between transmission peaks.  For the $z
< 4.5$\ ESI sample, the continuum was computed by both methods and is
plotted (diamonds) assuming the local interpolation.  See text for a
discussion of the comparison.  Where the transmitted flux is less than
0.0015, we have shown it at this nominal value.
}
\label{fig1a}
\addtolength{\baselineskip}{10pt}
%\end{inlinefigure}
\end{figure}

\newpage
%
% Figure 1b
%
\begin{figure}
\epsfig{figure=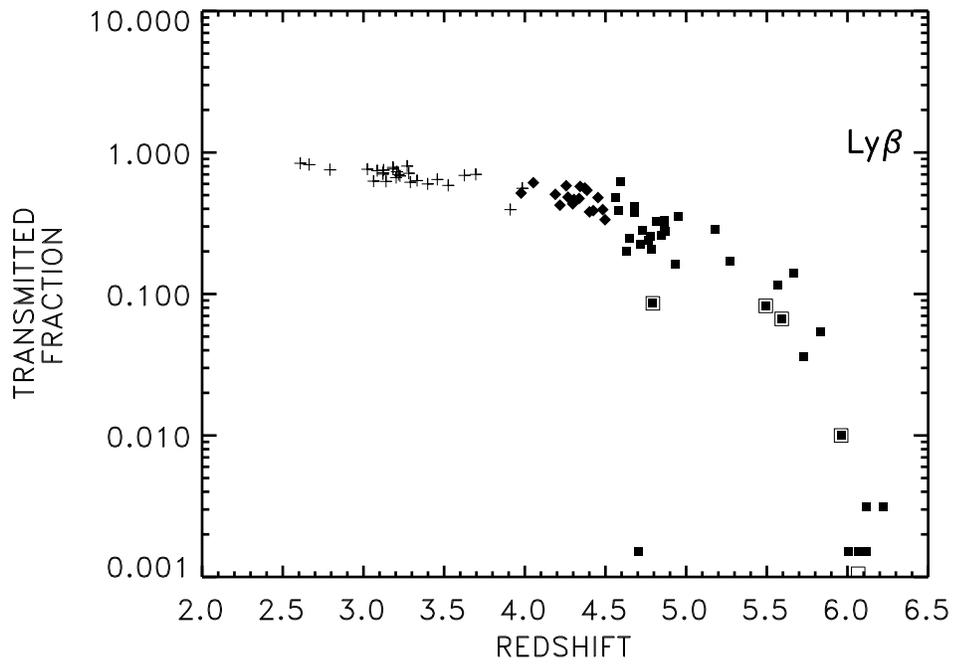,angle=0,width=5.0in}
\vspace{6pt}
\figurenum{1b}
\caption{As in Fig. 1a for transmission blueward of Ly$\beta$.  Flux
  was computed in two bins of width $15{\rm\AA}$\ between
  $980{\rm\AA}$\ and $1010{\rm\AA}$.  
}
\label{fig1b}
\addtolength{\baselineskip}{10pt}
\end{figure}

\newpage
%
% Figure 2a
%
\begin{figure}
\epsfig{figure=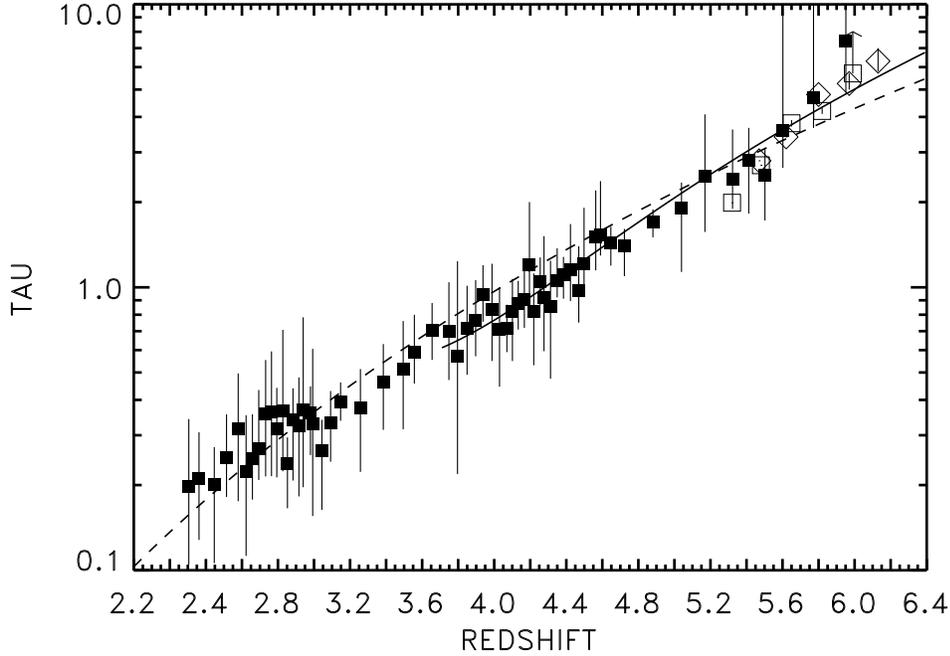,angle=0,width=5.0in}
\vspace{6pt}
\figurenum{2a}
\caption{Optical depth blueward of Ly$\alpha$\ emission as a
function of redshift for the complete sample of Figure~1a.  Starting
with the highest redshift points, mean transmissions were computed in
bins of six data points.  The optical depths corresponding to the mean
transmissions are plotted (filled squares) as a function of the median
redshift in each bin.  The error bars show the range of extremes of
transmission within each bin, converted to optical depth.  Mean
transmissions and median redshifts in each bin are tabulated in Table
2a, along with the redshift range in each bin.  The dashed
line shows a uniform ionized Gunn-Peterson model, normalized to match
the observations.  The solid line is the model of Equation~(2), with
$g$\ set equal to 
$0.74 \times ((1+z)/6)^{-4.1}$.  
Open symbols denote Ly$\alpha$\ optical depths from
Figure~6 of White et al.\ (2003) for the quasars SDSS 1148+5251
(diamonds) and SDSS 1030+0524 (squares).
}
\label{fig2a}
\addtolength{\baselineskip}{10pt}
\end{figure}

\newpage
%
% Figure 2b
%
\begin{figure}
\epsfig{figure=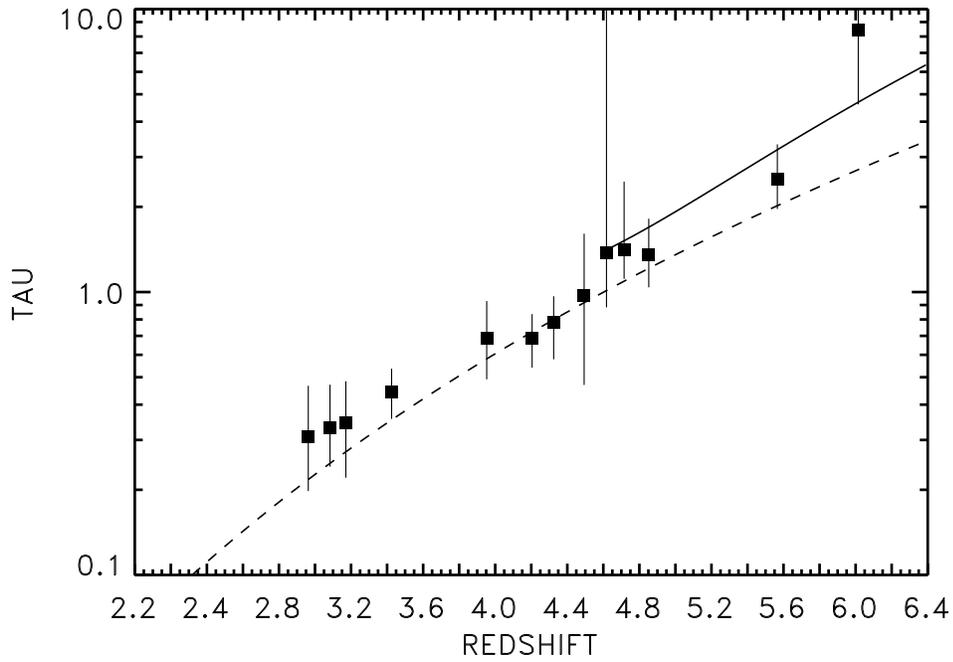,angle=0,width=5.0in}
\vspace{6pt}
\figurenum{2b}
\caption{As in Fig. aA for Ly~$\beta$.  The dashed line is a
  uniform ionized Gunn Peterson model  for Ly~$\beta$,
  computed with the same normalization as in Figure~2a.  It 
  underpredicts the Ly~$\beta$\ absorption because of the inhomogeneous
  nature of the IGM.  The solid line is the computed absorption from
  Equation~3 for the same run of ionization as in Figure~2a.  See
  text for discussion.
}
\label{fig2b}
\addtolength{\baselineskip}{10pt}
\end{figure}

\newpage
%Figure 3

\begin{figure}
\psfig{figure=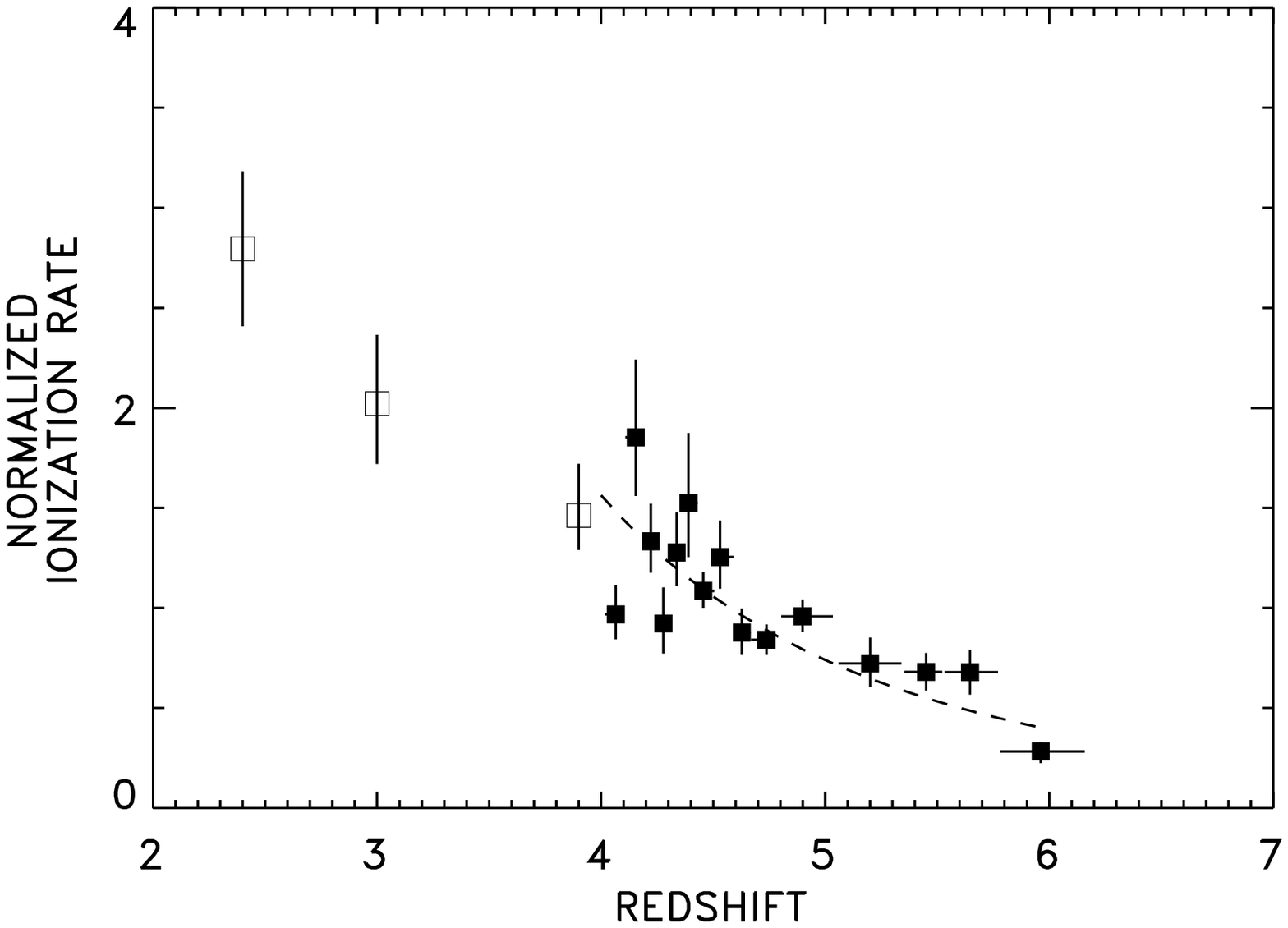,angle=0,width=5.5in}
\vspace{6pt}
\figurenum{3}
\caption{Evolution of the normalized ionization rate as a function of
redshift computed from mean transmissions using Equation~2 ({\it
filled squares}).  Mean transmissions and dispersions and median
redshifts were computed in 12-point bins, starting at the highest
redshift.  The horizontal error bars show the range of redshift in
each bin and the vertical error bars show the ionization parameter
corresponding to $\pm 1~\sigma$\ in the mean transmission in each bin.
The open squares are the measurements of McDonald et al.\
(2000) at lower redshift.  The dashed line is 
a power law of the form $0.74 \times ((1+z)/6)^{-4.1}$, fitted to the
present data for $z > 4$.
%a straight line fit to
%these lower redshift data and to the present data for $z > 4.7$.  
}
\label{fig3}
\addtolength{\baselineskip}{10pt}
\end{figure}

\newpage
%Figure 4

\begin{figure}
\psfig{figure=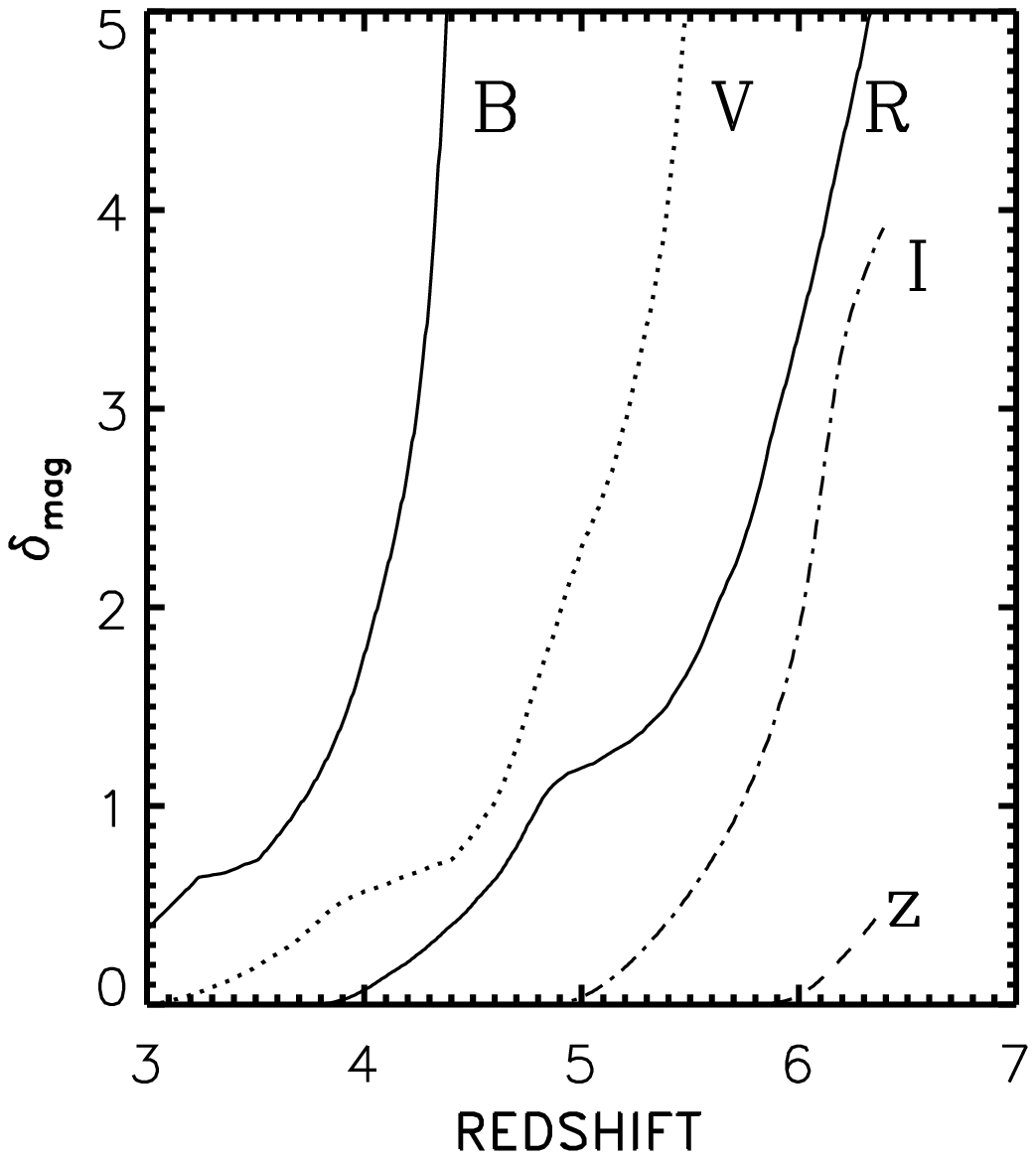,angle=0,width=5.0in}
\vspace{6pt}
\figurenum{4}
\caption{
Magnitude change expected as a result of IGM opacity as a function of
redshift for a flat $f_{\nu}$\ spectrum galaxy with a complete break at
the Lyman continuum edge, for each of 5 colors: Johnson $B$\ and $V$,
Kron-Cousins $R$\ and $I$\ and Sloan $z$.  The Suprime-Cam filter
transmissions (http://www.noaj.org/Observing/Instruments/SCam/sensitivity.html) 
were used for the specific
transmissions adopted.  Approximate forest
decrements were computed from the Ly$\alpha$, Ly$\beta$, and
Ly$\gamma$\ transmissions of Table~2.
}
\label{fig4}
\addtolength{\baselineskip}{10pt}
\end{figure}

\end{document}